\documentclass[1p]{elsarticle}
\usepackage{graphicx} 
\usepackage{amsmath}
\usepackage{psfrag}

\begin{document}
\title{An efficient, second order method for the approximation of the Basset history force}

\author[AP,M]{M.A.T. van Hinsberg}
\author[M]{J.H.M. ten Thije Boonkkamp}
\author[AP,MT]{ H.J.H. Clercx$^{*,}$}
\address[AP]{Department of Physics, Eindhoven University of Technology, PO Box 513,
 5600MB Eindhoven, The Netherlands}
\address[M]{Department of Mathematics and Computer Science, Eindhoven University of Technology, PO Box 513,
 5600MB Eindhoven, The Netherlands}
\address[MT]{Department of Applied Mathematics, University of Twente, PO Box 217, 7500 AE Enschede, The Netherlands}
\begin{abstract}
The hydrodynamic forces exerted by a fluid on small isolated rigid spherical particles are usually well described by the Maxey-Riley (MR) equation. The most time-consuming contribution in the MR equation is the Basset history force which is a well-known problem for many-particle simulations in turbulence. In this paper a novel numerical approach is proposed for the computation of the Basset history force based on the use of exponential functions to approximate the tail of the Basset force kernel. Typically, this approach not only decreases the cpu time and memory requirements for the Basset force computation by more than an order of magnitude, but also increases the accuracy by an order of magnitude. The method has a temporal accuracy of $\mathcal{O}\left(\Delta t^2\right)$ which is a substantial improvement compared to methods available in the literature. Furthermore, the method is partially implicit in order to increase stability of the computation. Traditional methods for the calculation of the Basset history force can influence statistical properties of the particles in isotropic turbulence, which is due to the error made by approximating the Basset force and the limited number of particles that can be tracked with classical methods. The new method turns out to provide more reliable statistical data.

\end{abstract}

\maketitle

\let\oldthefootnote\thefootnote
\renewcommand{\thefootnote}{\fnsymbol{footnote}}
\footnotetext[1]{Corresponding author. Tel: +31.40.247.2680, Fax: +31.40.246.4151, email: h.j.h.clercx@tue.nl}
\let\thefootnote\oldthefootnote

\section{Introduction}
 \noindent The turbulent dispersion of small inertial particles plays an important role in environmental flows, and in this work we focus on small particles with densities of the same order as that of the surrounding fluid. Examples of such particles that may be present in well-mixed or in density stratified estuaries are plankton, algae, aggregates (all with densities similar to the fluid density) or resuspended sand from the sea bottom (particle densities in this case several times that of the fluid). Particle collisions and the formation of aggregates of marine particles or sediment depend on the details of the small-scale trajectories of the particles in locally homogeneous and isotropic turbulence. At these scales the details of the hydrodynamic forces acting on (light) inertial particles are relevant.

Maxey and Riley~\cite{Maxey} introduced the equation of motion for small ($d_p\ll \eta$, with $d_p$ the particle diameter and $\eta$ the Kolmogorov length scale) isolated rigid spherical particles in a non-uniform velocity field ${\bf{u}}({\bf{x}},t)$. An important assumption is that the particle Reynolds number $Re_p=d_p|{\bf{u}}-{\bf{u}}_p|/\nu\ll 1$, with ${\bf{u}}_p$ the velocity of the particle and $\nu$ the kinematic viscosity of the fluid. The relative importance of the hydrodynamic forces depends on the ratio of particle-to-fluid density and the particle diameter. The computation of all the different forces in the Maxey-Riley equation is an expensive time- and memory consuming job. Therefore, assumptions are often made regarding the forces that can be neglected in the study of particle dispersion. The number of studies underpinning these assumptions, however, is rather limited due, for example, to lack of efficient algorithms to take into account the effects of the Basset history force with sufficient numerical accuracy. An elaborate overview of the work on the different terms in the Maxey-Riley equation and their numerical implementation can be found in the paper by Loth~\cite{loth00}.

The term most often neglected is the Basset history force because of its numerical complexity. Many recent studies underline the importance of the Basset force compared to the other hydrodynamic forces in the Maxey-Riley equation for particle transport in turbulent flows, see Refs.~\cite{mei91,Armenio,Marleen1,Marleen2}. Moreover, it can affect the motion of a sedimenting particle~\cite{sobral07} or bed-load sediment transport in open channels, where the Basset force becomes extremely important for sand particles \cite{Nino,Mordant}. It also might alter the particle velocity in an oscillating flow field~\cite{vojir94} or modify the trapping of particles in vortices~\cite{tanga94}.

Fast and accurate computation of the Basset force is far from trivial. Although several attempts have been made \cite{Bombardelli,Dorgan,Michaelides}, the computation of the Basset force is still far more time consuming and less accurate than the computation of the other forces in the MR equation. Therefore we present a new method that saves time, memory costs and is more accurate.

The MR equation and the subtlities with regard to the computation of the Basset history force are introduced in Section \ref{sec_part}. Next, in Section \ref{sec_tail} and \ref{sec_NA}, the new method is introduced, where Section \ref{sec_tail} focuses on the approximation of the tail of the Basset history force and Section \ref{sec_NA} on the numerical integration of the Basset history force. Thereafter, validation of the method using analytical solutions is discussed in Section \ref{sec_val}. A simulation of isotropic turbulence, with light inertial particles embedded in the flow, has been performed. In Section \ref{sec_iso} we compare the results from this simulation with the new implementation of the full MR equation with the old version used by van Aartrijk and Clercx~\cite{Marleen1}. Finally, concluding remarks are given in Section \ref{sec_con}.

\section{Particle tracking}\label{sec_part}
\noindent Particle trajectories in a Lagrangian frame of reference satisfy
\begin{eqnarray}
\frac{\textrm{d}\textbf{x}_p}{\textrm{dt}} =
\textbf{u}_p,\label{em1}
\end{eqnarray}
with $\textbf{x}_p$ the particle position and $\textbf{u}_p$ its
velocity. According to Maxey and Riley \cite{Maxey} the equation of
motion for an isolated rigid spherical particle in a nonuniform velocity field $\textbf{u}$
is given by

\begin{eqnarray}
m_p\frac{\textrm{d}\textbf{u}_p}{\textrm{dt}} &=& 6\pi a\mu
\left(\textbf{u}-\textbf{u}_p+\frac{1}{6}a^2\nabla^2\textbf{u}\right)
+m_f\frac{\textrm{D}\textbf{u}}{\textrm{Dt}} -(m_p-m_f)g\textbf{e}_z\nonumber\\
&&+\frac{1}{2}m_f\left(\frac{\textrm{D}\textbf{u}}{\textrm{Dt}}-\frac{\textrm{d}\textbf{u}_p}{\textrm{dt}}
+\frac{1}{10}a^2\frac{\textrm{d}}{\textrm{dt}}\left(\nabla^2\textbf{u}\right)\right)+6
a^2\rho\sqrt{\pi\nu}\int_{-\infty}^tK_{\textrm{B}}(t-\tau) \textbf{g}(\tau)\textrm{d}\tau\nonumber\\
&=&
\textbf{F}_{\textrm{St}}+\textbf{F}_{\textrm{P}}+\textbf{F}_{\textrm{G}}+\textbf{F}_{\textrm{AM}}+\textbf{F}_\textrm{B}.\label{em2}
\end{eqnarray}
The equation of motion includes time derivatives of the form
$\textrm{d}/\textrm{dt}$ taken along the particle path and
derivatives of the form $\textrm{D}/\textrm{Dt}$ taken along the
path of a fluid element. The particle mass is given by $m_p$, $a$ is
the radius of the particle, $\mu=\rho\nu$ is the dynamic viscosity,
$\rho$ and $\nu$ are the density of the fluid and its kinematic
viscosity, $m_f$ is the mass of the fluid element with a volume
equal to that of the particle and $\textbf{e}_z$ is the unit
vector in the opposite direction of the gravitational force. The forces in the
right-hand side of this equation denote the Stokes drag, local
pressure gradient in the undisturbed fluid, gravitational force,
added mass force and the Basset history force, respectively. The Fax\'{e}n correction proportional to $\nabla^2\textbf{u}$ has been included in the Stokes drag, added mass and Basset force \cite{faxen}. According to Homann \textsl{et al.} \cite{Homann} these corrections reproduce dominant finite-size effects on velocity and acceleration fluctuations for neutrally buoyant particles with diameter up to four times the Kolmogorov scale $\eta$. For the
added mass term the form described by Auton \textsl{et al.}
\cite{Auton} is used. Moreover, the history force convolution
function $\textbf{g}(t)$ and its kernel are
\begin{eqnarray}
\textbf{g}(t) = \frac{\textrm{d}\textbf{f}(t)}{\textrm{d}t}~,~~~~~~ \textbf{f}(t)=\textbf{u}-\textbf{u}_p+\frac{1}{6}a^2\nabla^2\textbf{u}~,~~~~~~
K_{\textrm{B}}(t)=\frac{1}{\sqrt{t}}. \label{em3}
\end{eqnarray}
Equation (\ref{em2}) is valid when $a\ll\eta$, but, as mentioned above, the Fax\'{e}n correction can weaken this condition. Furthermore, the particle Reynolds number must be small ($Re_p\ll 1$), as are the velocity gradients around the particle. Finally, the initial velocity of the particle and fluid must be equal. The coupled system (\ref{em1}) and (\ref{em2}) is in principle suitable for integration by any standard method, e.g. the fourth order Runge-Kutta method.

The Basset history force $\textbf{F}_\textrm{B}$ presents additional
challenges. First, the evaluation of the Basset force can become
extremely time consuming and memory demanding. This is due to the
fact that every time step an integral must be evaluated over the
complete history of the particle. Several attempts have been made to
solve this problem. Michaelides \cite{Michaelides} uses a Laplace
transform to find a novel way for computing the Basset force.
This procedure can be used for linear problems, but is not suitable
for space dependent velocity fields for which the coupled system
(\ref{em1}) and (\ref{em2}) is nonlinear. Another solution is
provided by Dorgan and Loth \cite{Dorgan} and Bombardelli \textsl{et
al.} \cite{Bombardelli}. In these papers the integral is evaluated
over a finite window from $t-t_{\textrm{win}}$ until $t$. This can
be represented by a change in the kernel of the Basset force. The window kernel is thus defined as
\begin{eqnarray}
K_{\textrm{win}}(t)=\left\{\begin{array}{ccc}
K_{\textrm{B}}(t)&\text{for}& t \leq t_{\textrm{win}},\\
0&\text{for}&t>t_{\textrm{win}}.\end{array}\right.\label{eq-window}
\end{eqnarray}
The kernel of the Basset
force is decreasing very slowly for $t\rightarrow \infty$, thus
$t_{\textrm{win}}$ must be chosen rather large. For Bombardelli
\textsl{et al.} \cite{Bombardelli} this problem turned out to be less important because they used a different kernel, which decreases faster for $t\rightarrow \infty$. Although the application of the window kernel saves CPU time, the computation of the Basset force is still far more expensive than the evaluation of the other forces in the MR equation. It turns out to be approximately 100 to 1000 times more time consuming depending on the application.

A second issue concerns the kernel of the Basset force, which is singular for $t\rightarrow0$. A standard approach to deal with the
singularity of the Basset kernel is to employ specific quadrature rules
such as the second order Euler-Maclaurin formula \cite{Press}.
Another approach is presented by Tatom \cite{Tatom} who uses a
fractional derivative method. This approach was tested by
Bombardelli \textsl{et al.} \cite{Bombardelli}. From their results
it can be easily shown that the integration method with specific
quadrature rules has only temporal accuracy $\mathcal{O}(\sqrt{\Delta t})$ and that the fractional
derivative approach has a temporal accuracy $\mathcal{O}(\Delta t)$. In computations of turbulent flows
with particles, other discretization methods involved are at least second order. Therefore, it is not sufficient to have a first order integration method for the Basset force.

Our goal is to derive a robust and efficient method for the computation of the Basset force that overcomes all the problems
mentioned above and to find an approach that is suitable for
different forms of the kernel. Furthermore, our method must be stable
and at least second order accurate in time. A third requirement is that it should be less
time consuming and memory demanding than previous methods.

\section{Approximation of the tail of the Basset force}\label{sec_tail}
\noindent To get a better understanding of the Basset force we will first show
that the contribution of this force is finite at any given time. To
do this, some restrictions on $\textbf{f}(t)$ and $\textbf{g}(t)=\frac{\textrm{d}}{\textrm{dt}}\textbf{f}(t)$ should be made. First, $\textbf{f}(t)$ must be a continuous function and its derivative must exist almost everywhere. Further, $\textbf{f}(t)$ and
$\textbf{g}(t)$ must be in the $\textrm{L}^{\infty}$space with norm $B_1$ and $B_2$, respectively. The restrictions on $\textbf{f}(t)$ and $\textbf{g}(t)$ are thus:
\begin{eqnarray}
\textbf{f}\in C^0, ~~~~~\|\textbf{f}\|_{\infty}=B_1,
~~~~~\|\textbf{g}\|_{\infty}=B_2,\label{cond1}
\end{eqnarray}
where $\|\cdot\|_{\infty}$ is defined as:
\begin{eqnarray}
\|\textbf{f}\|_{\infty}= \inf \{ C\ge 0:|\textbf{f}(t)| \le C \mbox{
almost everywhere}\},\label{cond2}
\end{eqnarray}
and $|\cdot|$ is the usual length of the vector. We assume that for particles in (turbulent) flows with $\textbf{f}(t)=\textbf{u}-\textbf{u}_p+\frac{1}{6}a^2\nabla^2\textbf{u}$ these conditions are satisfied as both the flow field and its Laplacian satisfy these conditions. With
the conditions in (\ref{cond1}) it is possible to find an upper bound for $\textbf{F}_B$.
The integral is split into two parts, in order to control both the
singularity in the Basset kernel and the tail of the integral. This yields
\begin{eqnarray}
\left|\frac{\textbf{F}_\textrm{B}}{c_\textrm{B}}\right|&=&
\left| \int_{-\infty}^tK_{\textrm{B}}(t-\tau)
\textbf{g}(\tau)\textrm{d}\tau
\right|\nonumber\\
&=&\left|
\int_{-\infty}^{t-\frac{B_1}{B_2}}\frac{\textbf{g}(\tau)}{\sqrt{t-\tau}}
\textrm{d}\tau+\int_{t-\frac{B_1}{B_2}}^t\frac{\textbf{g}(\tau)}{\sqrt{t-\tau}}\textrm{d}\tau
\right|\nonumber\\
&\leq&\left|\left[ \frac{
\textbf{f}(\tau)}{\sqrt{t-\tau}}\right]_{-\infty}^{t-\frac{B_1}{B_2}}-\int_{-\infty}^{t-\frac{B_1}{B_2}}\frac{
\textbf{f}(\tau)}{2(t-\tau)^{3/2}}\textrm{d}\tau\right|+\int_{t-\frac{B_1}{B_2}}^t\frac{\left|\textbf{g}(\tau)\right|}{\sqrt{t-\tau}}\textrm{d}\tau
\nonumber\\
&\leq&\sqrt{B_1B_2}+\frac{B_1}{2}\int_{-\infty}^{t-\frac{B_1}{B_2}}\frac{
1}{(t-\tau)^{3/2}}\textrm{d}\tau+B_2\int_{t-\frac{B_1}{B_2}}^t\frac{1}{\sqrt{t-\tau}}\textrm{d}\tau
\nonumber\\
&=& 4\sqrt{B_1B_2}.
\end{eqnarray}
Here $c_\textrm{B}=6
a^2\rho\sqrt{\pi\nu}$ is introduced for convenience. We now consider the window kernel for calculation of the Basset force $\textbf{F}_{\textrm{B-win}}$. In the limit of
$t_{\textrm{win}}\rightarrow\infty$ the difference between
$\textbf{F}_\textrm{B}$ and $\textbf{F}_{\textrm{B-win}}$ must
vanish. Using integration by parts, one can derive
\begin{eqnarray}
\left|\frac{\textbf{F}_\textrm{B}-\textbf{F}_{\textrm{B-win}}}{c_\textrm{B}}\right|&=& \left|
\int_{-\infty}^tK_{\textrm{B}}(t-\tau)\textbf{g}(\tau)\textrm{d}\tau-\int_{-\infty}^tK_{\textrm{win}}(t-\tau)\textbf{g}(\tau)\textrm{d}\tau
\right|\nonumber\\
&=&\left|
\int_{-\infty}^{t-t_{\textrm{win}}}\frac{\textbf{g}(\tau)}{\sqrt{t-\tau}}\textrm{d}\tau\right|\leq\frac{2B_1}{\sqrt{t_{\textrm{win}}}}.\label{win}
\end{eqnarray}
The error made by using the window kernel instead of the Basset
kernel is indeed becoming negligibly small for $t_{\textrm{win}}\rightarrow\infty$.
Unfortunately, this convergence is very slow, implying that $t_{\textrm{win}}$ must be very large, and a better approach for the computation of the Basset force
must be found. This is done by introducing a new kernel with a modified tail, in short the modified Basset kernel $K_{\textrm{mod}}(t)$, as follows
\begin{eqnarray}
&&K_{\textrm{mod}}(t)=\left\{\begin{array}{ccc}
K_{\textrm{B}}(t)&\textrm{for}& t \leq t_{\textrm{win}}\\
K_{\textrm{tail}}(t)&\textrm{for}&t>t_{\textrm{win}}\end{array}\right.\nonumber\\
&&\lim_{t\rightarrow\infty} K_{\textrm{tail}}(t)=0.
\label{kmod}
\end{eqnarray}
 This new kernel also implies a modified history force denoted by $\textbf{F}_{\textrm{B-mod}}$. For now $K_{\textrm{tail}}(t)$ is not yet defined but must be chosen such as to approximate the Basset kernel as close as possible. Using integration by parts in the last step, the upper bound for the error induced by the modified Basset force $\textbf{F}_{\textrm{B-mod}}$ becomes:
\begin{eqnarray}
\left|\frac{\textbf{F}_\textrm{B}-\textbf{F}_{\textrm{B-mod}}}{c_\textrm{B}}\right|&=& \left|
\int_{-\infty}^tK_{\textrm{B}}(t-\tau)\textbf{g}(\tau)\textrm{d}\tau-\int_{-\infty}^tK_{\textrm{mod}}(t-\tau)\textbf{g}(\tau)\textrm{d}\tau
\right|\nonumber\\
&=&\left|
\int_{-\infty}^{t-t_{\textrm{win}}}(K_{\textrm{B}}-K_{\textrm{tail}})(t-\tau)\textbf{g}(\tau)\textrm{d}\tau\right|\nonumber\\
&\leq&
B_1\left\{\Big|K_{\textrm{B}}(t_{\textrm{win}})-K_{\textrm{tail}}(t_{\textrm{win}})\Big|+\int_{t_{\textrm{win}}}^\infty\left|\frac{\textrm{d}
(K_{\textrm{B}}-K_{\textrm{tail}})(t)}{\textrm{dt}}\right|\textrm{d}t\right\}.\label{mod}
\end{eqnarray}
As the upper bound in relation (\ref{mod}) depends on $t_\textrm{win}$, it turns out to be beneficial to rescale the time and kernel as follows:
\begin{eqnarray}
\widetilde{K}_{\textrm{tail}}(\widetilde{t})=\frac{K_{\textrm{tail}}(t)}{K_{\textrm{B}}(t_{\textrm{win}})},~~~~~\widetilde{t}=\frac{t}{t_{\textrm{win}}}.\label{trans}
\end{eqnarray}
Applying the same scaling to ${K}_{\textrm{B}}(t)=1/\sqrt{t}$ we find
\begin{eqnarray}
\widetilde{K}_{\textrm{B}}(\widetilde{t})=\frac{K_{\textrm{B}}(t)}{K_{\textrm{B}}(t_{\textrm{win}})}={K}_{\textrm{B}}(\widetilde{t}).
\end{eqnarray}
Note that this cannot be done for a general kernel. Eq. (\ref{mod}) can now be reformulated as
\begin{eqnarray}
\left|\frac{\textbf{F}_\textrm{B}-\textbf{F}_{\textrm{B-mod}}}{c_\textrm{B}}\right|\leq
\frac{B_1}{\sqrt{t_{\textrm{win}}}}\left\{\left|1-\widetilde{K}_{\textrm{tail}}(1)\right|+\int_{1}^\infty\left|\frac{\textrm{d}
(K_{\textrm{B}}-\widetilde{K}_{\textrm{tail}})(\widetilde{t})}{\textrm{d}\widetilde{t}}\right|\textrm{d}\widetilde{t}\right\}.\label{min}
\end{eqnarray}
When comparing (\ref{win}) and (\ref{min}) one can see that a good approximation $\widetilde{K}_{\textrm{tail}}(\widetilde{t})$ of the tail reduces the error in (\ref{min}) significantly in comparison with (\ref{win}).

In order to find a good approximation $\widetilde{K}_{\textrm{tail}}(\widetilde{t})$ we start with (\ref{mod}).
The right hand side of (\ref{mod}) can be minimized and
thereby minimizing the error in $\textbf{F}_{\textrm{B-mod}}$. When determining
$K_{\textrm{tail}}(t)$ it is important that computation time is kept low. In order to achieve this, exponential functions are used because they can be implemented in a recursive way as explained later on. At first we start with one exponential function as follows,
\begin{eqnarray}
K_{\textrm{tail}}(t)=a\exp\left(-bt\right).
\end{eqnarray}
Here $a$ and $b$ are two positive constants. As a first guess we require that $K_{\textrm{tail}}(t_{\textrm{win}})=K_{\textrm{B}}(t_{\textrm{win}})$ and $\frac{\textrm{d}}{\textrm{d}t}K_{\textrm{tail}}(t_{\textrm{win}})=\frac{\textrm{d}}{\textrm{d}t}K_{\textrm{B}}(t_{\textrm{win}})$ in order to determine $a$ and $b$. In this way $K_{\textrm{mod}}(t)$, defined in (\ref{kmod}), is continuously differentiable. Doing this results in
\begin{eqnarray}
K_{\textrm{tail}}(t)=\sqrt{\frac{e}{t_{\textrm{win}}}}\exp\left(-\frac{t}{2t_{\textrm{win}}}\right).\label{eq-ab}
\end{eqnarray}
\begin{figure}[!hbtp]
\centering
\includegraphics[width=8cm]{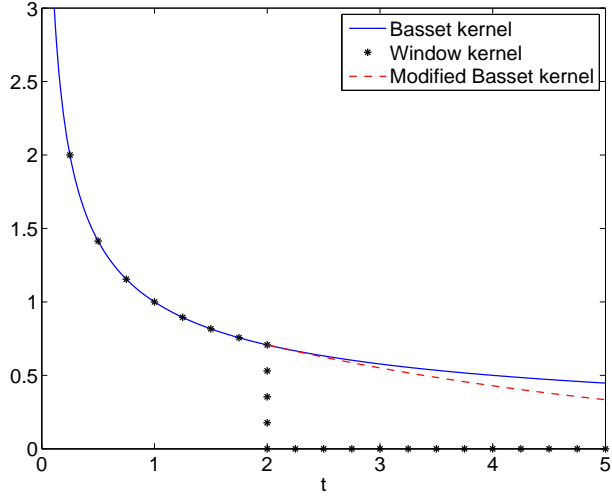}
 \caption{ Basset kernel (solid line), window kernel (dots) and the modified Basset kernel (dashed line) for $t_{\textrm{win}}=2$. }\label{fig-kernel}
\end{figure}
Fig. \ref{fig-kernel} shows several kernels, where the modified Basset kernel is given by (\ref{eq-ab}).
The error by applying the modified Basset kernel is obviously smaller compared to the error for the window method. In order to minimize the error even more, multiple exponential functions can be used. Relation (\ref{eq-ab}) provides an ansatz for the choice of $a$ and $b$. Thus we write $K_{\textrm{tail}}(t)$ as
\begin{eqnarray}
K_{\textrm{tail}}(t)=\sum_{i=1}^m a_iK_i(t),~~~~~K_i(t)=\sqrt{\frac{e}{t_i}}\exp\left(-\frac{t}{2t_i}\right),\label{eq-tail}
\end{eqnarray}
 with $a_i$ and $t_i$ positive constants. The functions $K_i(t)$ satisfy the following properties: $K_i(t_i)=K_\textrm{B}(t_i)$ and $\frac{\textrm{d}}{\textrm{d}t}K_i(t_i)=\frac{\textrm{d}}{\textrm{d}t}K_\textrm{B}(t_i)$. Combining (\ref{trans}) and (\ref{eq-tail}), we obtain the following dimensionless representation for the tail:
\begin{eqnarray}
\widetilde{K}_{\textrm{tail}}(\widetilde{t})=\sum_{i=1}^m a_i\widetilde{K}_{i}(\widetilde{t})~~,~~ \widetilde{K}_{i}(\widetilde{t})=\sqrt{\frac{e}{\widetilde{t}_i}}\exp\left(-\frac{\widetilde{t}}{2\widetilde{t}_i}\right)~~,~~
\widetilde{t}_i=\frac{t_i}{t_{\textrm{win}}}.\label{tail}
\end{eqnarray}
The coefficients $a_i$ and $\widetilde{t}_i$ should be chosen in such a way that the upper bound in (\ref{min}) is minimized.
However, Newton iteration will not work for this problem, and instead we consider the expression
\begin{eqnarray}
\left(1-\widetilde{K}_{\textrm{tail}}(1)\right)^2+\int_{1}^\infty
\widetilde{t} \left(\frac{\textrm{d}
(K_{\textrm{B}}-\widetilde{K}_{\textrm{tail}})}{\textrm{d}\widetilde{t}}\right)^2\textrm{d}\widetilde{t},\label{min2}~,
\end{eqnarray}
which provides a good indication for the optimal values of $a_i$ and $\widetilde{t}_i$. In (\ref{min2}) an extra multiplication with $\widetilde{t}$ is introduced to correct for
the change in norm. After minimizing the expression in (\ref{min2}), we can verify whether the error in
(\ref{min}) is of the same order.
Since $\widetilde{K}_i(\widetilde{t}_i)=K_\textrm{B}(\widetilde{t}_i)$ and $\frac{\textrm{d}}{\textrm{d}\widetilde{t}}\widetilde{K}_i(\widetilde{t}_i)=\frac{\textrm{d}}{\textrm{d}\widetilde{t}}K_\textrm{B}(\widetilde{t}_i)$, the function $\widetilde{K}_i(\widetilde{t})$ approximate $K_\textrm{B}(\widetilde{t})$ very well around $\widetilde{t}_i$. The kernel $K_{\textrm{B}}$ must be approximated over a large range of $\widetilde{t}$-values and as a consequence $\widetilde{t}_i$ must also have a large range. Furthermore, $K_{\textrm{B}}$ is
changing slowly for large $\widetilde{t}$ so the small $\widetilde{t}_i$ must be close to
each other whereas the large $\widetilde{t}_i$ can be far apart. The approach for finding $a_i$ and $\widetilde{t}_i$ is thus the following. First, make a reasonable choice for $\widetilde{t}_i$, and second, calculate $a_i$ by minimizing (\ref{min2}). Finally, determine the term
between brackets from (\ref{min}). Another slightly different set of $\widetilde{t}_i$-values can be chosen to see if a better approximation can be made. In Table \ref{table-1} the result is shown for
$m=10$. Here one can see that some values of $\widetilde{t}_i$ are smaller than 1. This is surprising because the kernel $K_{\textrm{B}}$ is not being approximated below $\widetilde{t}=1$. When tuning the $\widetilde{t}_i$-values we found, however, that this improves the approximation.
\begin{table}
\caption{Coefficients $a_i$ and $\widetilde{t}_i$ in $\widetilde{K}_{\textrm{tail}}(\widetilde{t})$ with $m=10$}
\centering
\begin{tabular}{ l c }
\hline\hline\noalign{\smallskip}
\vspace{0.1cm}
$\widetilde{t}_i$ & $a_i$ \\

  \hline
  0.1 & 0.23477481312586  \\
  0.3 & 0.28549576238194 \\
  1 &  0.28479416718255 \\
  3 &  0.26149775537574 \\
  10 & 0.32056200511938 \\
  40 & 0.35354490689146 \\
  190 & 0.39635904496921\\
  1000 &  0.42253908596514\\
  6500 &   0.48317384225265\\
  50000 &  0.63661146557001 \\
  \hline
\end{tabular}
\label{table-1}
\end{table}

\begin{figure}
\centering
\begin{minipage}[t]{0.48\linewidth}
\centering
\psfrag{K}{\scriptsize $K_{\textrm{B}}$}
\psfrag{AApp}{\scriptsize $\widetilde{K}_{\textrm{tail}}$}
\psfrag{t}{\scriptsize $\widetilde{t}$}
\includegraphics[width=6.6cm]{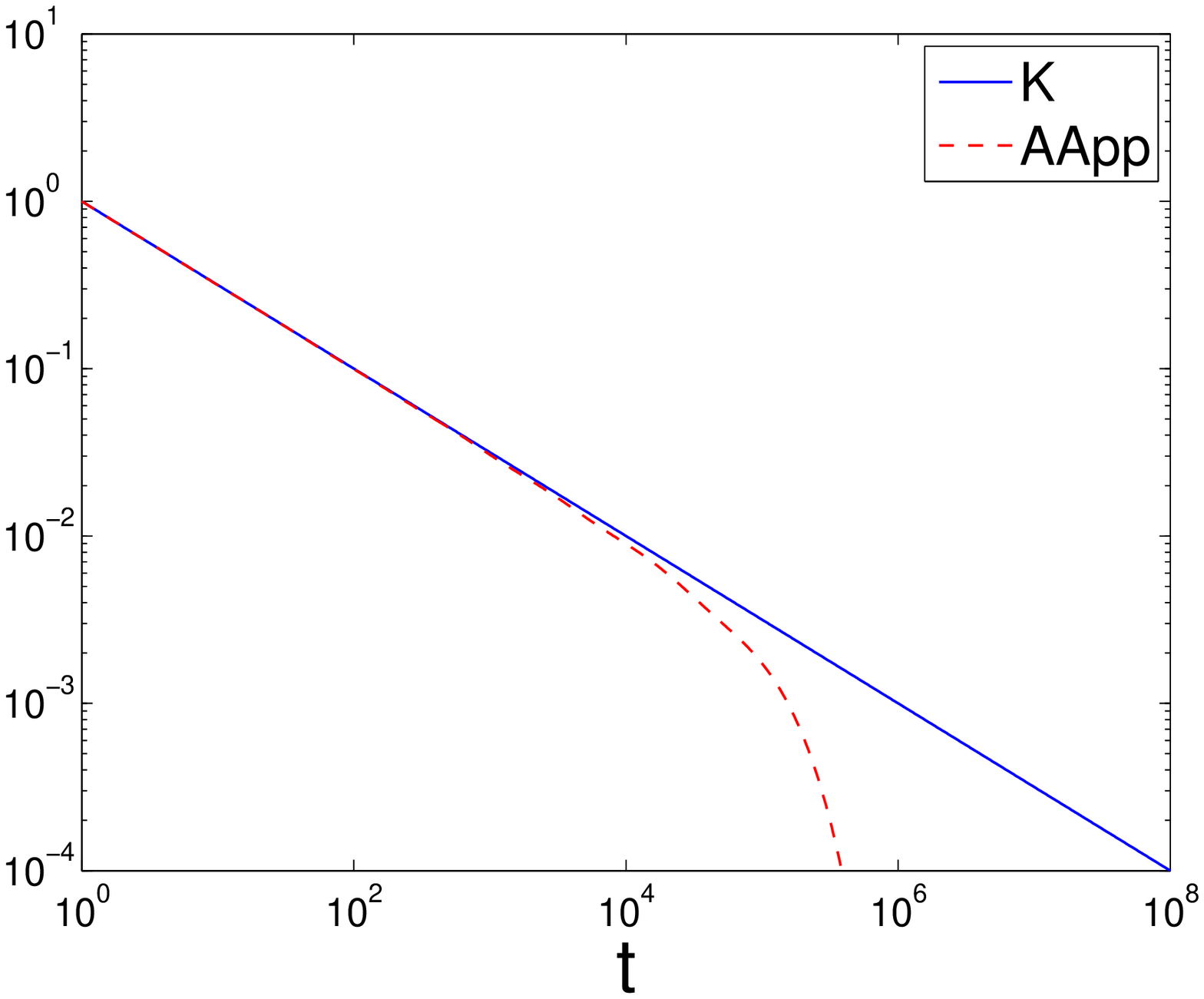}
\caption{The kernels $K_{\textrm{B}}(\widetilde{t})$ and $\widetilde{K}_{\textrm{tail}}(\widetilde{t})$.}
\label{fig1}
\end{minipage}%
\hspace{0.5cm}%
\begin{minipage}[t]{0.48\linewidth}
\centering
\psfrag{AAApppp}{\scriptsize $\left|K_{\textrm{B}}-\widetilde{K}_{\textrm{tail}}\right|$}
\psfrag{t}{\scriptsize $\widetilde{t}$}
\includegraphics[width=6.6cm]{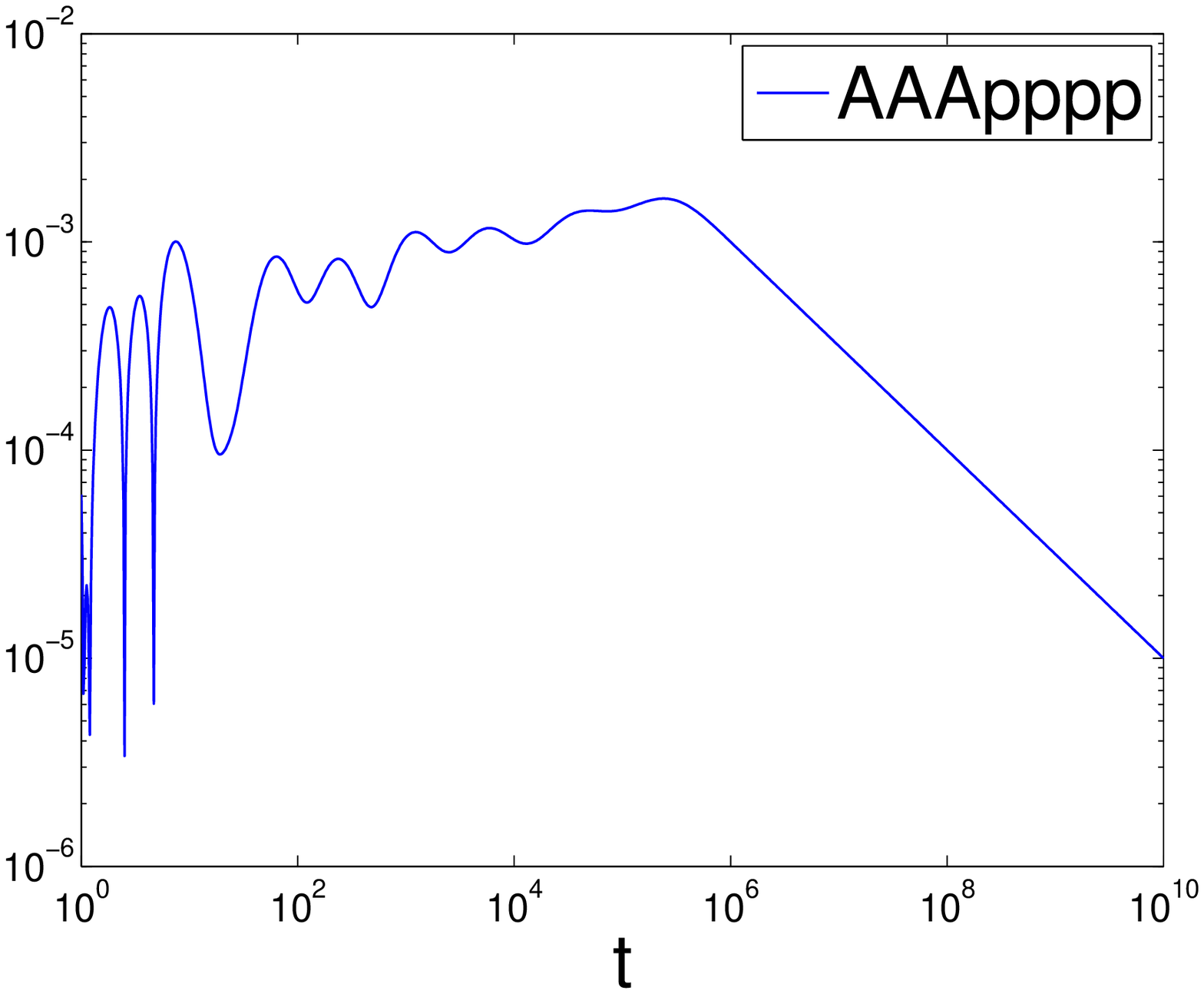}\caption{The error $\left|K_{\textrm{B}}-\widetilde{K}_{\textrm{tail}}\right|$.}\label{fig2}
\end{minipage}
\end{figure}

From Fig. \ref{fig1} it can be seen that $\widetilde{K}_{\textrm{tail}}$ approximates $K_{\textrm{B}}$ relatively well over a wide range of $\widetilde{t}$. From Fig. \ref{fig2} one can see that the error decays for large $\widetilde{t}$ (note the huge range of $\widetilde{t}$ in both figures).

Using (\ref{tail}) in combination with Table \ref{table-1} for $\widetilde{K}_{\textrm{tail}}(\widetilde{t})$ the part between brackets in (\ref{min}) can be calculated
\begin{eqnarray}
\left|1-\widetilde{K}_{\textrm{tail}}(1)\right|+\int_{1}^\infty\left|\frac{\textrm{d}
(K_{\textrm{B}}-\widetilde{K}_{\textrm{tail}})(\widetilde{t})}{\textrm{d}\widetilde{t}}\right|\textrm{d}\widetilde{t}\approx9.5\cdot10^{-3}.
\end{eqnarray}
Comparing this result with the window method (\ref{win}) a factor of more than 200 is
gained in accuracy. When keeping the same accuracy but changing
the window, $t_{\textrm{win}}$ can be decreased by a factor of  $200^2=40000$.
\section{Numerical approximation}\label{sec_NA}
\noindent In this section the numerical integration is discussed. First, the integration of the window and tail kernels are elaborated. Second, the overall numerical scheme for solving Eq. (\ref{em1}) and (\ref{em2}) is explained.

The integration of the Basset force with the modified kernel (\ref{kmod}) and (\ref{eq-tail}) is split into two parts, the window kernel and the tail of the kernel as follows,
\begin{eqnarray}
\textbf{F}_{\textrm{B-mod}}(t)&=&
c_\textrm{B}\int_{-\infty}^tK_{\textrm{mod}}(t-\tau)\textbf{g}(\tau)\textrm{d}\tau\nonumber\\
&=&c_\textrm{B}\int_{t-t_\textrm{win}}^tK_{\textrm{B}}(t-\tau)\textbf{g}(\tau)\textrm{d}\tau+c_\textrm{B}\int_{-\infty}^{t-t_\textrm{win}}K_{\textrm{tail}}(t-\tau)\textbf{g}(\tau)\textrm{d}\tau\nonumber\\
&=&\textbf{F}_{\textrm{B-win}}(t)+\textbf{F}_{\textrm{B-tail}}(t).
\end{eqnarray}
In the following, methods are described for the calculation of $\textbf{F}_{\textrm{B-win}}$ and $\textbf{F}_{\textrm{B-tail}}$.

First, we consider the Basset force due to the window kernel $\textbf{F}_{\textrm{B-win}}$. The kernel of the Basset force is singular for $t\rightarrow0$ which impedes use of the ordinary trapezoidal rule. In order to deal with the singularity we introduce an alternative, trapezoidal-based method, referred to as the TB-method. The idea is as follows. The trapezoidal rule is based on linear interpolation of the integrand on each subinterval. In our approach $\textbf{g}(t)$ is approximated by its linear interpolant $\textbf{P}_1(t)$, and subsequently the integration of $K_{\textrm{B}}(t-\tau)\textbf{P}_1(\tau)$ is done exactly.
For the numerical implementation we start with the discretization of the interval $[t-t_\textrm{win},t]$, given by $\tau_n=t-n\Delta t,~~n=0,1,2,\cdots,N$ with $\Delta t=t_\textrm{win}/N$. Now the integral can be split as
\begin{equation}
\textbf{F}_{\textrm{B-win}}(t)=c_\textrm{B}\sum_{n=1}^{N}\int_{\tau_n}^{\tau_{n-1}} \frac{\textbf{g}(\tau)}{\sqrt{t-\tau}}\textrm{d}\tau~.
\end{equation}
The next step is to approximate $\textbf{g}(\tau)$ by its linear interpolant on each subinterval, which yields
\begin{equation}
\textbf{F}_{\textrm{B-win}}(t)\approx c_\textrm{B}\sum_{n=1}^{N}\int_{\tau_n}^{\tau_{n-1}} \frac
{\textbf{g}_n+(\textbf{g}_{n-1}-\textbf{g}_{n})(\tau-\tau_n)/\Delta t}{\sqrt{t-\tau}}\textrm{d}\tau,
\end{equation}
where $\textbf{g}_n\equiv \textbf{g}(\tau_n)$. After the change of variable $\tau'=t-\tau$ this integral can be evaluated and the following result\footnote{This formulation is preferred to avoid loss of significant digits in the computation of $\textbf{F}_{\textrm{B-win}}$.} is obtained:
\begin{eqnarray}
\textbf{F}_{\textrm{B-win}}(t)  \approx \frac{4}{3}c_\textrm{B} \textbf{g}_0\sqrt{\Delta
t}+c_\textrm{B}\textbf{g}_N\frac{\sqrt{\Delta
t}\left(N-\frac{4}{3}\right)}{(N-1)\sqrt{N-1}+(N-\frac{3}{2})\sqrt{N}}\nonumber\\
+c_\textrm{B}\sqrt{\Delta
t}\sum_{n=1}^{N-1} \textbf{g}_n
\left(\frac{n+\frac{4}{3}}{(n+1)\sqrt{n+1}+
(n+\frac{3}{2})\sqrt{n}}+\frac{n-\frac{4}{3}}{(n-1)\sqrt{n-1}+(n-\frac{3}{2})\sqrt{n}}\right).\label{coef}
\end{eqnarray}
From the result above one can see that three inner products must be calculated each time step, one inner product for each spatial dimension. One vector contains all the values $\textbf{g}_n$ which must be shifted by one index each time step. The other vector containing the coefficients in (\ref{coef}) is calculated once at the start of the computation. In this way the computational time is kept minimal. The part with $\textbf{g}_0$ will be treated in a different way as explained later on in order to improve stability.

Next, the numerical integration of the tail of the Basset force is discussed. The idea is to find a recursive formulation in order to minimize computation efforts. Using expression (\ref{eq-tail}) for $K_{\textrm{tail}}$, $\textbf{F}_{\textrm{B-tail}}$ becomes:
\begin{eqnarray}
\textbf{F}_{\textrm{B-tail}}(t)= \sum_{i=1}^ma_ic_\textrm{B}\int_{-\infty}^{t-t_\textrm{win}} K_i\left(t-\tau\right)\textbf{g}(\tau)\textrm{d}\tau=\sum_{i=1}^m a_i\textbf{F}_i(t)~,
\end{eqnarray}
Here, $\textbf{F}_i$ represents the contribution of the $i$-th exponential function. Now $\textbf{F}_i$ is split into two parts, as follows.
\begin{multline}
\textbf{F}_i(t)=c_\textrm{B}\int_{t-t_\textrm{win}-\Delta t}^{t-t_\textrm{win}} K_i(t-\tau)
\textbf{g}(\tau)\textrm{d}\tau
+c_\textrm{B}\int_{-\infty}^{t-t_\textrm{win}-\Delta t} K_i(t-\tau)
\textbf{g}(\tau)\textrm{d}\tau\\=\textbf{F}_{i\textrm{-di}}(t)+\textbf{F}_{i\textrm{-re}}(t)~,
\end{multline}
where we have to compute $\textbf{F}_{i\textrm{-di}}$ directly and where $\textbf{F}_{i\textrm{-re}}$ can be computed recursively. For $\textbf{F}_{\textrm{i-di}}$ the same procedure is followed as with the window kernel. Using this procedure the following result\footnote{Note that in equation (\ref{direct}) Taylor series must be used for $\varphi\left(-\frac{\Delta t}{2t_i}\right)$ when $\Delta t\ll t_i$.} can be obtained:
\begin{multline}
\textbf{F}_{i\textrm{-di}}(t)\approx c_\textrm{B}\sqrt{\frac{e}{t_i}}\int_{t_\textrm{win}}^{t_\textrm{win}+\Delta t} \exp\left(-\frac{\tau'}{2t_i}\right)\left(\textbf{g}_{N}+\frac{t_\textrm{win}-\tau'}{\Delta t}(\textbf{g}_N-\textbf{g}_{N+1})\right)\textrm{d}\tau'
=2c_\textrm{B}\sqrt{et_i}\\\exp\left(-\frac{t_\textrm{win}}{2t_i}\right) \left\{ \textbf{g}_N\left[1-\varphi\left(-\frac{\Delta t}{2t_i}\right) \right]+\textbf{g}_{N+1}\exp\left(-\frac{\Delta t}{2t_i}\right)\left[\varphi\left(\frac{\Delta t}{2t_i}\right)-1\right]\right\} ,\label{direct}
\end{multline}
where $\varphi(z) = (e^z-1)/z = 1+\frac{1}{2}z+\frac{1}{6}z^2+\mathcal{O}\left(z^3\right)$.
Finally, $\textbf{F}_{\textrm{i-re}}$ can be easily calculated using the value of $\textbf{F}_i$ at the previous time step:
\begin{eqnarray}
\textbf{F}_{i\textrm{-re}}(t)&=&c_\textrm{B}\int_{-\infty}^{t-t_\textrm{win}-\Delta t} \sqrt{\frac{e}{t_i}}\exp\left(-\frac{t-\tau}{2t_i}\right)
\textbf{g}(\tau)\textrm{d}\tau\nonumber\\
&=&\exp\left(-\frac{\Delta t}{2t_i}\right)c_\textrm{B}\int_{-\infty}^{t-t_\textrm{win}-\Delta t} \sqrt{\frac{e}{t_i}}\exp\left(-\frac{t-\Delta t-\tau}{2t_i}\right)
\textbf{g}(\tau)\textrm{d}\tau \nonumber\\
&=&\exp\left(-\frac{\Delta t}{2t_i}\right)\textbf{F}_{\textrm{i}}(t-\Delta t)~.
\end{eqnarray}

In this last part the overall numerical scheme is discussed. To solve equation (\ref{em1}) and (\ref{em2}) numerically the second-order Adams-Bashforth (AB2) method is implemented. For a differential equation $\frac{\textrm{d}\textbf{y}}{\textrm{d}t}=\textbf{h}(t,\textbf{y})$ the scheme reads $\textbf{y}_{n+1}=\textbf{y}_n+\frac{\Delta t}{2}\left(3\textbf{h}^n-\textbf{h}^{n-1}\right)$, where $\textbf{h}^n=\textbf{h}(t^n,\textbf{y}^n)$. Equation (\ref{em1}) can be directly integrated with this scheme but for equation (\ref{em2}) some modifications are needed. In order to have a stable scheme, the $\frac{\textrm{d}\textbf{u}_p}{\textrm{dt}}$  term in the added mass force is treated in an implicit way instead of explicit. Moreover, it turned out that the AB2-method has poor stability properties for the calculation of the Basset force using the window method. Extremely small time steps must be taken in order to have a stable solution. An alternative method circumventing stability problems is to bring a part of the Basset force (the contribution  $\frac{\textrm{d}\textbf{u}_p}{\textrm{dt}}$ evaluated at $t$) to the left hand side. Eq. (\ref{em2}) is then reformulated as
\begin{eqnarray}
\left(m_p+\frac{1}{2}m_f+\frac{4}{3}c_\textrm{B} \sqrt{\Delta t}\right)\frac{\textrm{d}\textbf{u}_p}{\textrm{dt}} =
\textbf{F}_{\textrm{St}}+\textbf{F}_{\textrm{P}}+\textbf{F}_{\textrm{G}}+\textbf{F}_{\textrm{AM}}'+\textbf{F}_\textrm{B}'~,\label{eq-impl}
\end{eqnarray}
with $\textbf{F}_{\textrm{AM}}'=\frac{1}{2}m_f\left(\frac{\textrm{D}\textbf{u}}{\textrm{Dt}}+\frac{1}{10}a^2\frac{\textrm{d}}{\textrm{dt}}(\nabla^2\textbf{u})\right)$ and $\textbf{F}_{\textrm{B}}'=\textbf{F}_{\textrm{B}} - \frac{4}{3}c_\textrm{B} \sqrt{\Delta t} \frac{\textrm{d}\textbf{u}_p}{\textrm{dt}}$. In this way the Basset force becomes partially implicit instead of completely explicit. Finally, as only the time derivative along the particle path $\frac{\textrm{d}\textbf{u}}{\textrm{dt}}$ is available, the time derivative along the path of a fluid element $\frac{\textrm{D}\textbf{u}}{\textrm{Dt}}$ is computed according to
\begin{eqnarray}
\frac{\textrm{D}\textbf{u}}{\textrm{Dt}}=\frac{\partial\textbf{u}}{\partial t}+u_j\frac{\partial\textbf{u}}{\partial x_j}=\frac{\partial\textbf{u}}{\partial t}+u_{p,j}\frac{\partial\textbf{u}}{\partial x_j}+(u_j-u_{p,j})\frac{\partial\textbf{u}}{\partial x_j}=\frac{\textrm{d}\textbf{u}}{\textrm{dt}}+(u_j-u_{p,j})\frac{\partial\textbf{u}}{\partial x_j}~.
\end{eqnarray}
\section{Validation of the Basset force integration}\label{sec_val}
\noindent In this section four test cases are presented in order to validate the methods for the integration of the Basset force. The first example tests the trapezoidal-based (TB) method and compares the results with the semi-derivative (SD) approach by Bombardelli \textsl{et al.} \cite{Bombardelli}. Example 2 and 3 test the the overall numerical scheme. Here both stability and convergence are tested for the explicit and the partially implicit TB-method. Finally, example 4 shows the efficiency of the Basset force using the tail kernel.

\paragraph{Example 1: Basset integral for a given convolution function}
\mbox{}\\
In order to demonstrate the advantages of the TB-method, the convergence of this method is compared with the SD-approach of Bombardelli \textsl{et al.} \cite{Bombardelli}. To that end the arbitrary test function $g(\tau)=\cos\tau$ is used. The exact Basset integral is given by
\begin{eqnarray} F_\textrm{B}(t)&=&c_\textrm{B}\int_0^t\frac{\cos\tau}{\sqrt{t-\tau}}\textrm{d}\tau=2c_\textrm{B}\int_0^{\sqrt{t}}\cos(t-\sigma^2)\textrm{d}\sigma\nonumber\\
&=&c_\textrm{B}\sqrt{2\pi}\left(C(\sqrt{2t/\pi})\cos t + S(\sqrt{2t/\pi})\sin t\right),
 \end{eqnarray}
 with $\sigma=\sqrt{t-\tau}$ and $C(t)$ and $S(t)$ the Fresnel cosine and sine functions \cite{Fresnel}, respectively.

\begin{figure}[!hbtp]
\centering
\includegraphics[width=8cm]{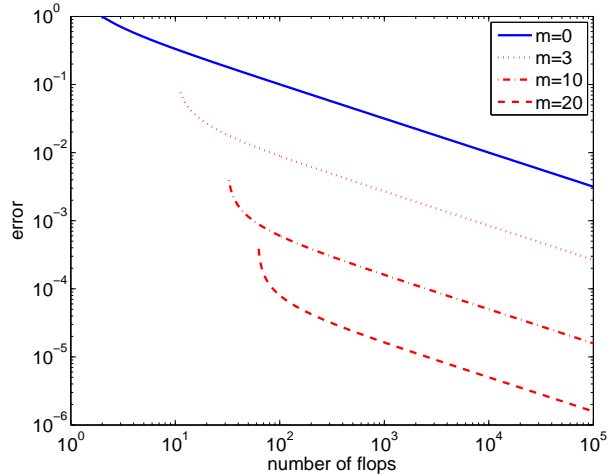}
 \caption{Upper bound for the error in the approximation of the Basset force as a function of the number of flops for different number of exponential functions (indicated by $m$).}\label{fig33}
\end{figure}

The Basset integral $F_\textrm{B}$ was evaluated at $t=50\pi$ with different numbers of points $N$ uniformly distributed in the interval $[0,50\pi]$. The results for both the SD-approach and the TB-method are presented in Table \ref{table-2}. Here, it can be seen that the error of the TB-method is substantially smaller than that of the SD-approach. When increasing the number of points $N$ it can be seen that the TB-method is second-order accurate in time (in agreement with analysis that can be done by using Taylor series), whereas the SD-approach is first-order accurate in time. More methods have been compared by Bombardelli \textsl{et al.} \cite{Bombardelli} but these methods have even lower order of convergence than the SD-approach.
 \begin{table}[!h]
\caption{Relative error and order of convergence for the Basset integral, for the SD-approach \cite{Bombardelli} and the TB-method.}
\centering
\begin{tabular}{ l c c c c}
\hline\hline
   & Relative error & Order & Relative error  & Order \\
  points $N$&SD &SD&TB  &TB \\
  \hline
  81 & $4.03\cdot10^{-1}$& &$1.34\cdot10^{-1}$  \\
  243 & $1.37\cdot10^{-1}$& 1.0&$2.54 \cdot10^{-2}$& 1.5\\
  729 & $4.66\cdot10^{-2}$&1.0  &$3.29\cdot10^{-3}$&1.9 \\
  2,187 &$1.56 \cdot10^{-2}$& 1.0& $3.93\cdot10^{-4}$&1.9\\
  6,561 &$5.22\cdot10^{-3}$&1.0& $4.54\cdot10^{-5}$&2.0 \\
  19,683 &$1.74\cdot10^{-3}$&1.0& $5.15\cdot10^{-6}$&2.0\\
  59,049 &$5.80\cdot10^{-4}$&1.0& $5.80\cdot10^{-7}$&2.0\\
  177,147 &$1.93\cdot10^{-4}$&1.0&  $6.49\cdot10^{-8}$&2.0\\
  531,441 &$6.45\cdot10^{-5}$&1.0&   $7.24\cdot10^{-9}$&2.0\\
  1,594,323 &$2.15\cdot10^{-5}$&1.0&  $8.06\cdot10^{-10}$ &2.0\\
  \hline
\end{tabular}
\label{table-2}
\end{table}
\paragraph{Example 2: Space-dependent steady velocity field}
\mbox{}\\
In order to test the overall numerical scheme for the computation of particle trajectories we have implemented a particular space-dependent steady velocity field. The particle trajectory is a circle and given by $(x(t),y(t))=(r\cos\omega t, -r\sin\omega t)$, where $r$ and $\omega$ denote the radius and the angular velocity, respectively. The velocity field and its derivation is given in appendix A. For the test case, exactly one revolution is simulated, from $t=0$ until $t=2\pi$. In order to test the stability of the overall scheme two different approaches have been tested. One with the completely explicit time integration procedure for the Basset force and the other with the partially implicit procedure, see Section \ref{sec_NA}. For both the implicit and explicit method the Basset force is computed with the TB-method and show second-order convergence in $\Delta t$. The relative error is computed with  $\textbf{x}_p(2\pi)$. The results are presented in Table \ref{table-3} and clearly indicate that the explicit scheme is very unstable when the number of time steps is smaller than 256. The partially implicit scheme remains stable even with the number of time steps as small as 16.
\begin{table}[!hbtp]
\caption{Relative error and order of convergence for the overall numerical scheme, tested for the trajectory of a small particle in a space dependent steady velocity field.}
\centering
\begin{tabular}{ l c c c c}
\hline\hline
   number of&  Relative error & Order &  Relative error & Order \\
   time steps &  explicit& explicit &  implicit & implicit \\

  \hline
  16 & unstable&&$ 3.63 \cdot10^{-1}$ \\
  32 &unstable&&$ 8.32 \cdot10^{-2}$& 2.1 \\
  64 &unstable&&$ 2.09 \cdot10^{-2}$& 2.0 \\
  128 &unstable&&$ 5.28 \cdot10^{-3}$& 2.0  \\
  256 &$ \textsl{4.80} \cdot\textsl{10}^{\textsl{-2}}$&&$ 1.33 \cdot10^{-3}$& 2.0 \\
  512 &$ 3.05 \cdot10^{-4}$& \textsl{7.3}&$ 3.33 \cdot10^{-4}$& 2.0 \\
  1024 &$ 7.68 \cdot10^{-5}$& 2.0&$ 8.34 \cdot10^{-5}$& 2.0  \\
  2048 &$ 1.93 \cdot10^{-5}$& 2.0&$ 2.09 \cdot10^{-5}$& 2.0  \\
  4096 &$ 4.84 \cdot10^{-6}$&2.0&$ 5.21 \cdot10^{-6}$&2.0 \\
  \hline
\end{tabular}
\label{table-3}
\end{table}
\paragraph{Example 3: Time-dependent velocity field}
\mbox{}\\
The trajectory of a spherical particle in an arbitrary time-dependent velocity field can rather straightforwardly be computed as long as the velocity field is smooth enough. The derivation of the particle trajectory uses Laplace transforms and the analytical procedure is given in appendix B. The overall numerical scheme is tested by computing the trajectory of a particle in the following one-dimensional, time-dependent velocity field
\begin{eqnarray}
u(t)=\frac{(m_p-m_f)g}{6\pi a\mu} \cos 2t~.\label{eq-vel_field}
\end{eqnarray}
The total force on the particle is zero at $t=0$, i.e, $\textbf{F}_{\textrm{St}}$ and $\textbf{F}_{\textrm{G}}$ are in balance. In order to compute the Basset force the implicit TB-method is used. The integration is carried out from $t=0$ until $t=2\pi$. The relative error is computed for $u_p(2\pi)$ and is presented in Table \ref{table-4}, where once again second-order time accuracy is confirmed.
From these test cases, using both a time-dependent and a space-dependent velocity field for the computation of particle trajectories, we can conclude that the (partially implicit) TB-method is stable and second-order accurate in time, and conjecture that this remains the case for particles in arbitrary time- and space-varying flow fields.
\begin{table}[!hbtp]
\caption{Relative error and order of convergence for the overall numerical scheme, for the velocity field (\ref{eq-vel_field}).}
\centering
\begin{tabular}{ l c c }
\hline\hline
   time steps &  Relative error & Order  \\

  \hline
  16 &$ 9.96 \cdot10^{-2}$&  \\
  32 &$ 2.38 \cdot10^{-2}$& 2.1 \\
  64 &$ 5.57 \cdot10^{-3}$& 2.1 \\
  128 &$ 1.31 \cdot10^{-3}$& 2.1 \\
  256 &$ 3.13 \cdot10^{-4}$& 2.1 \\
  512 &$ 7.56 \cdot10^{-5}$& 2.0 \\
  1024 &$ 1.84 \cdot10^{-5}$& 2.0 \\
  2048 &$ 4.53 \cdot10^{-6}$& 2.0\\
  4096 &$ 1.12 \cdot10^{-6}$& 2.0\\
  \hline
\end{tabular}
\label{table-4}
\end{table}
\paragraph{Example 4: Computational efficiency due to modified kernel integration}
\mbox{}\\
In this example the computational savings when using the modified tail kernel, given in (\ref{kmod}) and (\ref{eq-tail}), is investigated based on analysis of the number of flops per time step, per particle and per space dimension. For the window kernel this is $N+1$ flops because only one vector dot product is calculated. For each exponential function three extra flops are needed. To see how efficient the tail kernel works the upper bound (\ref{min}) for the error is plotted as a function of the computation time, Fig \ref{fig33}. Different numbers (indicated by $m$) of exponential functions are taken into account. The results are plotted in Fig \ref{fig33}. Here it can be seen that the best choice for $m$ depends on the particular situation. Furthermore, the results show a significant saving in computation time. This can easily be a factor of 100 or more. When looking to the memory requirements the results are even better. For the window method as many memory locations as flops are needed whereas each exponential function only takes one memory location instead of 3 flops. So using the tail kernel not only saves time but also memory.

Overall, the use of the tail kernel reduces the computational costs of the Basset force by more than an order of magnitude, whereas the memory requirement is even reduced more. Furthermore, the error is reduced by more than an order of magnitude. The question remains, of course, whether the computational savings directly result in faster simulations. This depends on the remaining part of the simulation. Although the other force contributions in (\ref{eq-impl}) can be calculated much faster than the Basset force this does not have to hold for the interpolation of the velocities in a turbulence simulation. The velocity of the flow field is only computed at the grid points and an interpolation must be carried out to compute the velocity at the particle position. This may be very time consuming and it can become the new bottleneck. The reduction of CPU-time might then not be as big as expected but it remains significant. Additionally, the decrease in memory requirement may become essential when increasing the number of particles in turbulence simulations.
\section{Light particles in isotropic turbulence}\label{sec_iso}
\noindent In this section a brief statistical analysis of velocities of particles, released in an isotropic turbulent flow, is provided. The isotropic turbulence simulation is performed by means of direct numerical simulations. The numerical code consist of two parts. First the Navier-Stokes equations with the Boussinesq approximation are solved on a triple periodic domain using a pseudo-spectral code \cite{Marleen-thesis,Marleen3} (Eulerian approach). Second, the particle trajectories are obtained by the Lagrangian approach as explained in the previous sections.
%
The simulation is performed on a $128^3$ grid. The number of (light) particles is 20,000 and the particle-to-fluid density ratio $\rho_p/\rho_f=4$ (thus all hydrodynamic forces in the MR equation are relevant, see Refs.~\cite{Marleen1,Marleen2}). The integral-scale Reynolds number is $\textrm{Re}=UL/\nu=1333$, with $U$ the typical root-mean-square velocity and $L$ the integral length scale. The Stokes number $\textrm{St}$ is typically in the range $0.1\le \textrm{St} \le 1.0$~\cite{Marleen2} and particles are tracked for a period of approximately two eddy turnover times.

Two simulations have been carried out under exactly the same flow conditions and particle tracking is either based on the classical approach (window method) or on the novel integration method (exponential method) for the Basset kernel. In the first simulation only the window kernel (\ref{eq-window}) has been used, where the number of time steps in the window is $n=500$. The other one uses the modified window kernel, given in (\ref{kmod}) and (\ref{eq-tail}). In this case only five time steps are taken into account in the window, so $n=5$. For the tail of the Basset kernel the number of exponential functions $m=10$.

\begin{figure}[!hbtp]
\centering
\begin{minipage}[t]{0.48\linewidth}
\centering
\psfrag{t}{$\tau$}
\includegraphics[width=6.6cm]{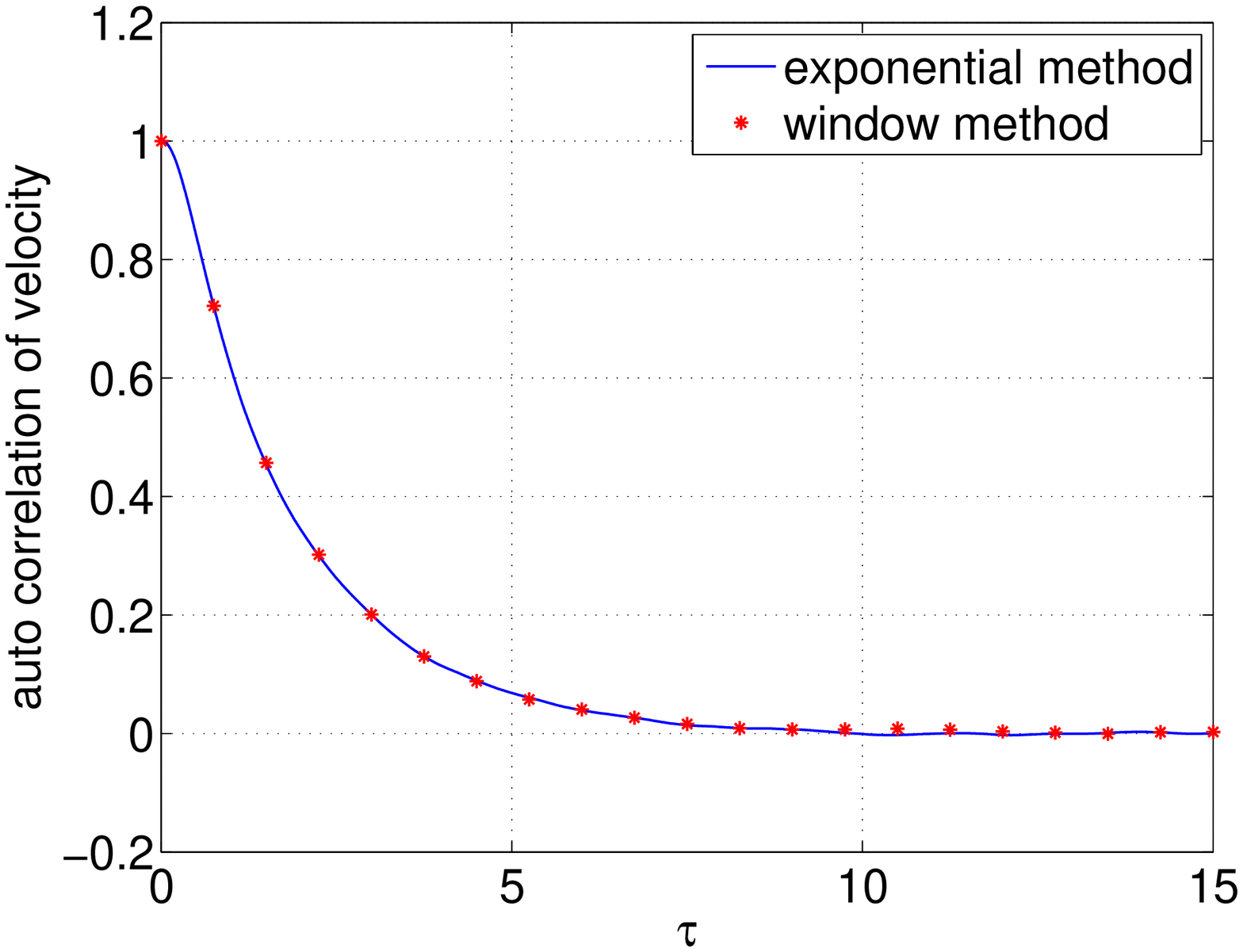}
\caption{Autocorrelation of the particle velocity ${\bf{u}}_p$. The solid line represents the result from the exponential method and the dots those from the window method.}\label{fig3}
\end{minipage}%
\hspace{0.5cm}%
\begin{minipage}[t]{0.48\linewidth}
\centering
\psfrag{w}{$\omega$}
\includegraphics[width=6.6cm]{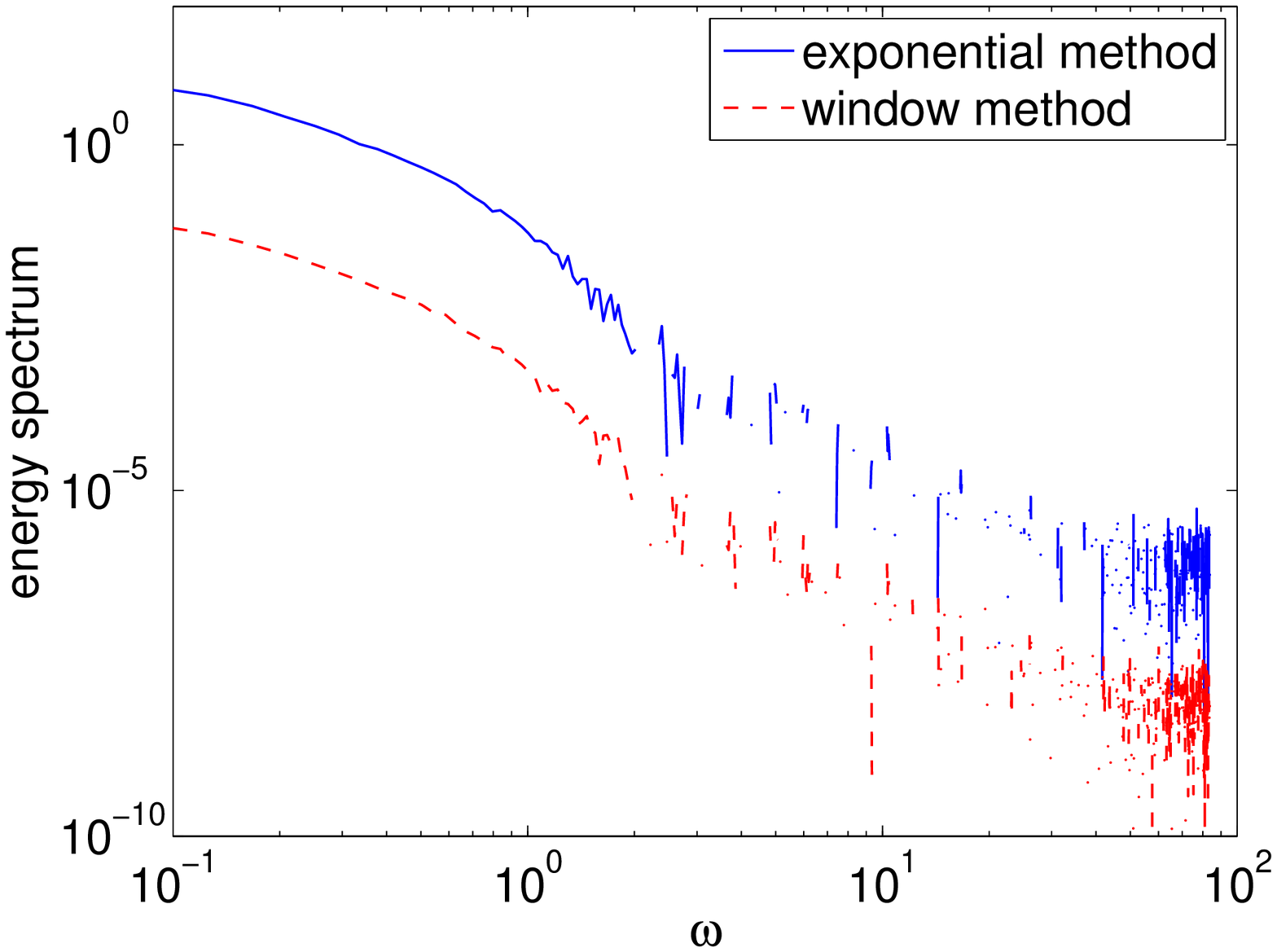}
\caption{Energy spectrum of the particle velocity. The graph of the window method (dashed line) is shifted downward with respect to the spectrum from the exponential method (solid line) by a factor of 100 for clarity.}\label{fig4}
\end{minipage}
\end{figure}
In order to study a particle trajectory we start with considering the energy spectrum of the particle. To obtain the energy spectrum, we first need to calculate the autocorrelation $R(\tau)$ of the velocity, which is defined by
\begin{eqnarray}
R(\tau)=\frac{\langle u_p(t)u_p(t+\tau)\rangle}{\langle u_p(t)^2\rangle}~.
\end{eqnarray}
Here, $\langle \cdot \rangle$ denotes the average in over the different particles. The particles are embedded in a homogeneous isotropic turbulent flow and no gravitation is applied. Therefore, we are allowed to average over the components of the velocity vector of all particles. No time averaging has been applied for the present velocity data as this run covers only one or two eddy turnover times. The results for the autocorrelation of the velocity are shown in Fig.\ref{fig3} and we see that the results for both the window method and the exponential method are comparable. The energy spectrum obtained from the particle velocities can be calculated by taking the cosine transform of the autocorrelation function, and is shown in Fig.\ref{fig4}. Although the results are similar we are interested in possible differences between the two spectra. If these differences have an overall trend this would mean that statistical properties can be influenced by the different methods of evaluating the Basset kernel. However, to observe any error in the evaluation of the Basset force kernel the differences should be larger than the statistical noise.

A starting point for an analytical evaluation of possible differences between the window method and the exponential method consists of the response of a single particle in a uniform oscillating flow field. We are therefore interested in the periodic solution $u_p$ of a spherical particle responding to an oscillating velocity field $u=\cos\omega t$ (or $u={\mathcal{R}}[\exp(i\omega t)]$, with $i$ the imaginary unit and ${\mathcal{R}}$ denoting the real part of this expression). The particle velocity can then be expressed as $u_p={\mathcal{R}}[V\exp(i\omega t)]$ with $V$ a complex amplitude, which is dependent on the method chosen to evaluate the Basset force kernel. For the window method and the exponential method we introduce  $V_\textrm{win}$ and $V_\textrm{exp}$, respectively. For $V_\textrm{exp}$ the exact solution $V_\textrm{ex}$ is used since the error of the exponential method is assumed to be negligibly small, see also Fig. \ref{fig2}. In general, $|V_\textrm{win}|\neq \left|V_\textrm{ex}\right|$ which means that some frequencies are suppressed with the window method while others may be amplified. This should become visible in the energy spectrum of particle velocities.

In order to find $V_\textrm{win}$ Eq. (\ref{em2}) should be solved for $u={\mathcal{R}}[\exp(i\omega t)]$ and $u_p={\mathcal{R}}[V\exp(i\omega t)]$, resulting in the following integro-differential equation:
\begin{multline}
i\omega m_pV_\textrm{win} = 6\pi a\mu \left(1-V_\textrm{win}\right)
+\frac{i\omega}{2}m_f \left(3-V_\textrm{win}\right)\\
+i\omega c_\textrm{B} (1-V_\textrm{win})\int_{t-t_\textrm{win}}^t\frac{e^{-i\omega (t-\tau)}}{\sqrt{t-\tau}}\textrm{d}\tau~.\label{eq-fresnel-1}
\end{multline}
Here, we used the fact that the velocity field is uniform, one dimensional and that no gravity is applied. Applying the change in variables $\sigma=\sqrt{(t-\tau)\omega}$, allows us to find an expression for $V_\textrm{win}$ i.e.,
\begin{eqnarray}
V_\textrm{win}&=& 1+\frac{(m_f-m_p)i\omega}{6\pi a\mu+\left(\frac{1}{2}m_f+m_p\right)i\omega +c_\textrm{B}\sqrt{2\omega\pi}Q(\sqrt{2t_\textrm{win}\omega\pi})}\label{eq-fresnel-3},
\end{eqnarray}
where $Q(t)=S(t)+iC(t)$, with $C(t)$ and $S(t)$ the Fresnel cosine and sine functions, respectively~\cite{Fresnel}.
$V_{\textrm{ex}}$ can now be found by taking $V_{\textrm{ex}}=\lim_{t_\textrm{win}\rightarrow\infty}V_\textrm{win}$ which results in
\begin{eqnarray}
V_\textrm{ex}&=& 1+\frac{(m_f-m_p)i\omega}{6\pi a\mu+\left(\frac{1}{2}m_f+m_p\right)i\omega +c_\textrm{B}\sqrt{\frac{\omega\pi}{2}}(1+i)}.
\end{eqnarray}
Inspection of the energy spectrum displayed in Fig.~\ref{fig4} reveals that the noise becomes more important for increasing $\omega$. The effects from the different methods for the computation of the Basset force kernel turned out to be most important for $\omega\geq1$. Unfortunately, the noise in the energy spectrum is already larger than predicted for the differences between the window and exponential method. One way of decreasing the error would be averaging over time, but with the limited number of eddy turnover times in the present simulation this is not feasible. However, an alternative approach exists in comparing the autocorrelation of the particle acceleration.
\begin{figure}
\centering
\begin{minipage}[t]{0.48\linewidth}
\centering
\psfrag{t}{$\tau$}
\includegraphics[width=6.6cm]{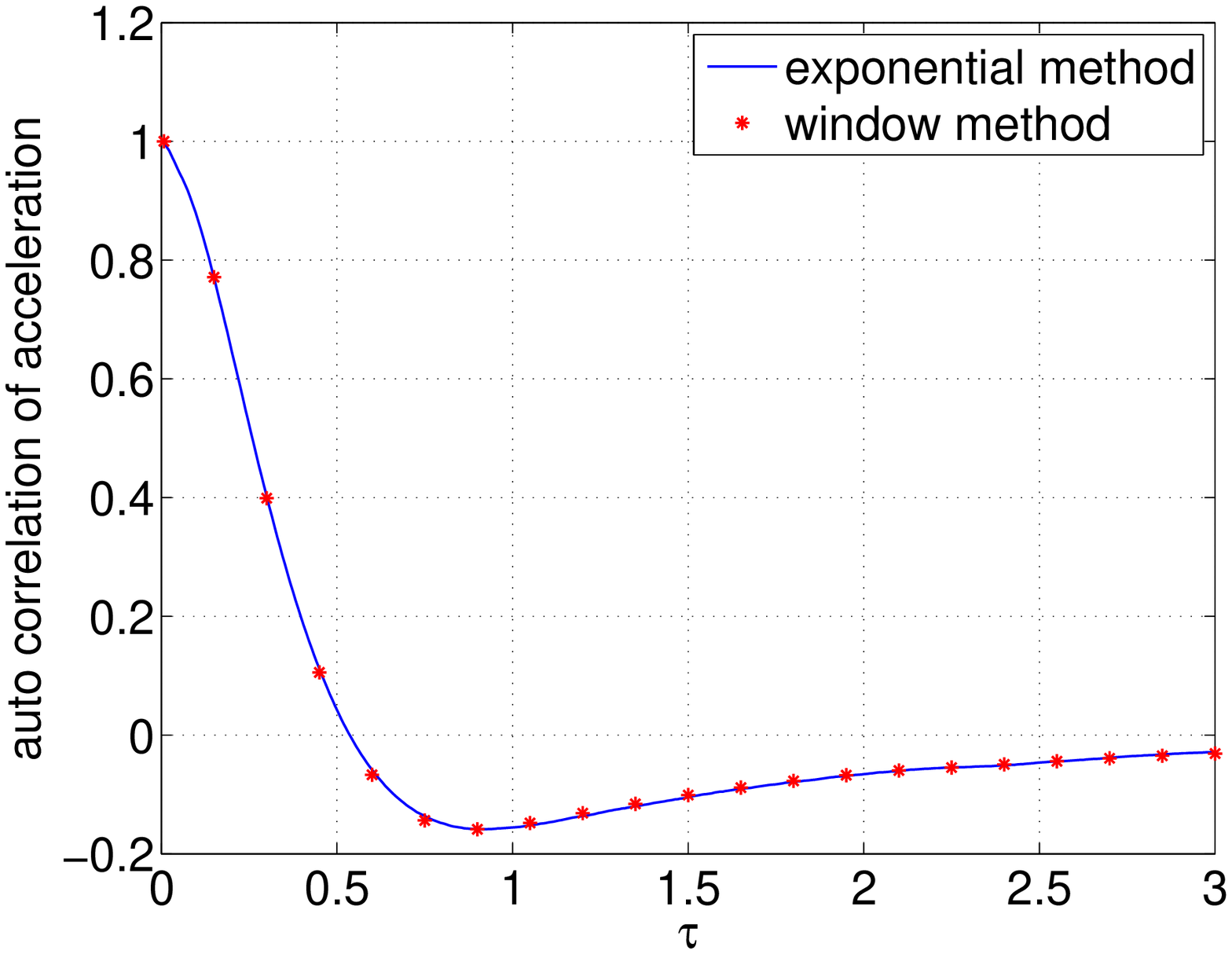}
\caption{Autocorrelation of the particle acceleration ${\bf{a}}_p=d{\bf{u}}_p/dt$. The solid line represents the result from the exponential method and the dots those from the window method.} \label{fig5}
\end{minipage}%
\hspace{0.5cm}%
\begin{minipage}[t]{0.48\linewidth}
\centering
\psfrag{w}{$\omega$}
\includegraphics[width=6.6cm]{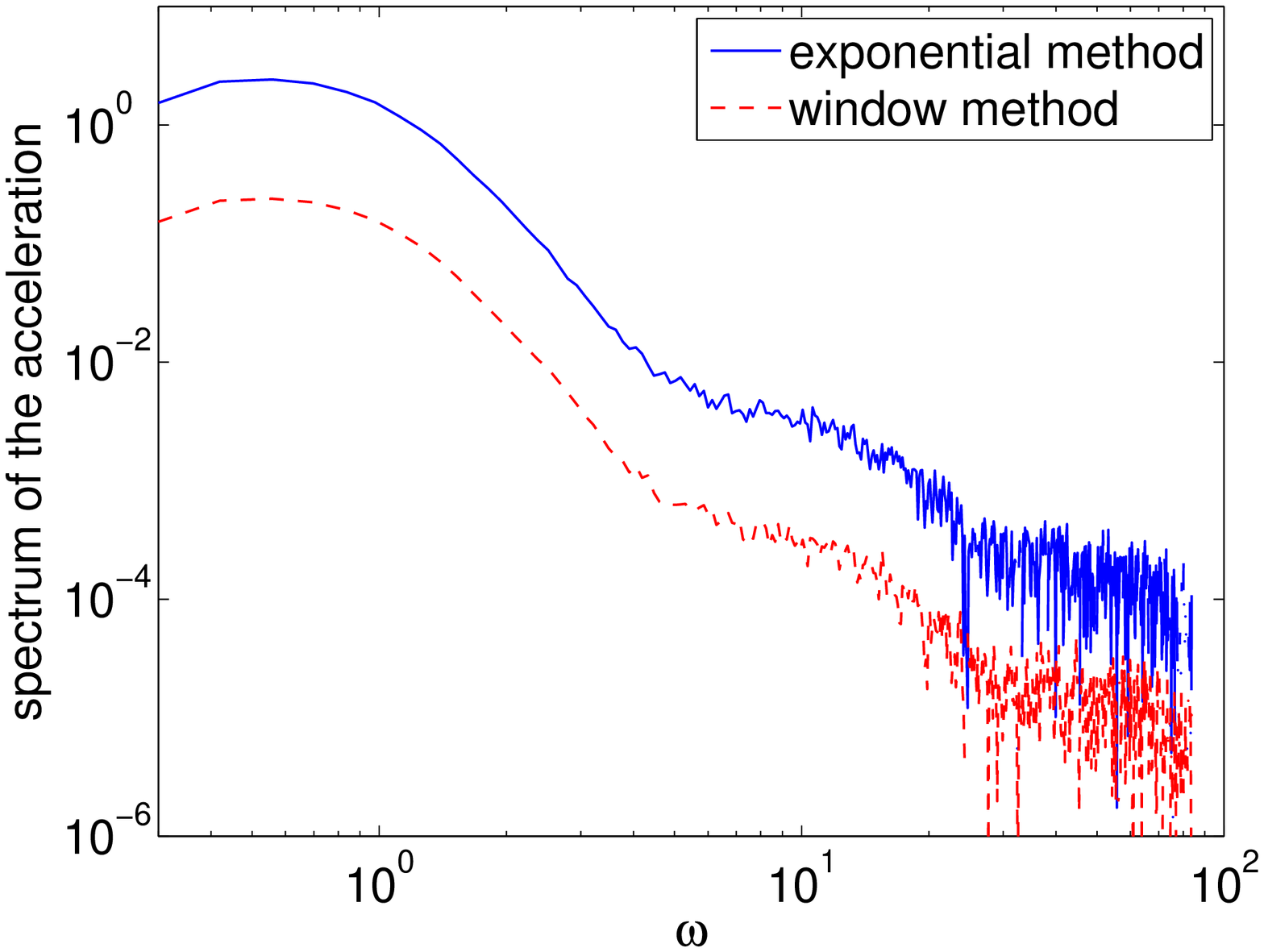}
\caption{Spectrum of the particle acceleration. The graph of the window method (dashed line) is shifted downward with respect to the spectrum from the exponential method (solid line) by a factor of 10 for clarity.} \label{fig6}
\end{minipage}
\end{figure}

\begin{figure}[!h]
\centering
\psfrag{w}{$\omega$}
\includegraphics[width=8cm]{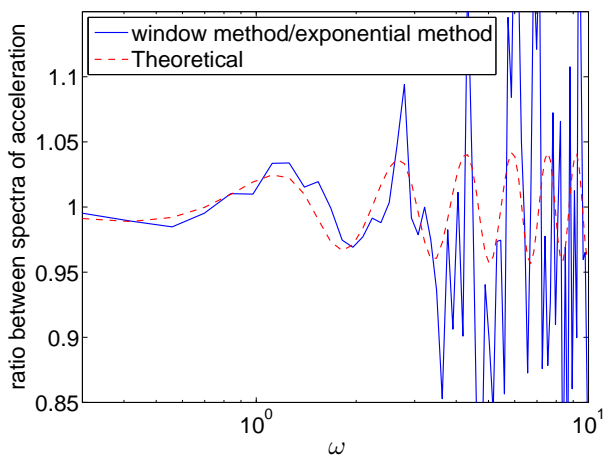}
\caption{The theoretical ratio $\left(\frac{|V_\textrm{win}|}{|V_\textrm{ex}|}\right)^2$ (dashed line) compared with a similar ratio of the particle acceleration spectra (solid line).}\label{fig7}
\end{figure}
The autocorrelation of the particle acceleration is plotted in Fig. \ref{fig5}. Here, the typical time scale is much shorter than that of the particle velocity, therefore it is possible to also average over time. The spectrum is calculated by taking the cosine transform of the autocorrelation acceleration function and is displayed in Fig. \ref{fig6}. Because the particle acceleration is used instead of the particle velocity, higher frequenties (shorter time scales) become more important. In this way deteriorating influence of the noise on the spectrum is shifted to higher frequenties. Nevertheless, the computed spectrum should still be affected by the method that is chosen for the evaluation of the Basset force kernel. In order to observe the differences the best approach is to plot the ratio of the two spectra as function of frequency. When no essential difference exists between the window and exponential method their ratio would be equal to one with some noise added to it. However, the window method suppresses some frequency components while others are amplified, so the deviation from one is a measure for the error in the window method. In Fig. \ref{fig7} the ratio of both spectra is shown in combination with the theoretical ratio defined by $\left(\frac{|V_\textrm{win}|}{|V_\textrm{ex}|}\right)^2$. 
From Fig. \ref{fig7} it can be seen that the theoretical ratio predicts the ratio obtained from the simulation, including the local maxima and minima quite well, provided the frequency is not too high. For higher frequencies the noise becomes larger but the theoretical and computational ratios still seem to have the same trend. The novel exponential method to evaluate the Basset force kernel might be considered as an excellent and efficient method for tracking of many particles in turbulent flows.

\section{Conclusions}\label{sec_con}
We have introduced a novel method for the evaluation of the Basset force kernel and analysed several aspects of its implementation. The tail of the Basset force kernel is approximated by exponential functions. The contribution of these exponential functions can be calculated in a recursive way which makes it very efficient. Typically the use of the tail kernel reduces the computational costs of the Basset force by more than an order of magnitude, whereas the memory requirement is reduced even more. Furthermore, the error in the tail of the Basset force is also reduced by more than an order of magnitude in comparison with the traditional window method.

A trapezoidal-based method is developed in order to deal with the singularity of the Basset force. This method has a temporal accuracy of $\mathcal{O}\left(\Delta t^2\right)$ where other methods only have a temporal accuracy of $\mathcal{O}\left(\Delta t\right)$ or lower. This method is made partially implicit in order to make it more stable.

The method has been implemented in a tracking algorithm for (light) inertial particles in turbulent flows.
The isotropic turbulence simulation shows that the error made by the window method can influence statistics on the particle trajectories. This has been illustrated with the velocity and acceleration spectra. Therefore, the novel exponential method is preferred over the classical window method. Because the new implementation is much faster than the classical one, more particles can be taken into account in simulations, which opens possibilities for further research.
\section{Acknowledgements}
We thank Ben van den Broek for contributing
to the derivation of the analytical solution presented in Appendix B.
\appendix
\numberwithin{equation}{section}
\section{Flow field for circular particle trajectories}
\noindent In this appendix the space dependent velocity field is derived that allows a circular particle trajectory as solution of the MR equation. Suppose the particle trajectory and velocity is given by
\begin{eqnarray}
\textbf{x}_p=r{\mathcal{R}}\left[\textbf{e}~e^{-i\omega t}\right],~~~~\textbf{u}_p=-r \omega{\mathcal{R}}\left[i\textbf{e}~e^{-i\omega t}\right]~,\label{eq-x_p}
\end{eqnarray}
with $\textbf{e}=\textbf{e}_x-i\textbf{e}_y$. For the flow velocity field we are looking for solutions of (\ref{em2}) of the form
\begin{eqnarray}
\textbf{u}=-{\mathcal{R}}\left[s(x+iy)\textbf{e}\right] + 2\alpha z\textbf{e}_z ~,\label{eq-u}
\end{eqnarray}
with $s=\alpha+i\beta$ a complex constant. For $\beta=0$ the velocity field represents a sink flow $\textbf{u}=(-\alpha x, -\alpha y, 2\alpha z)$ and for $\alpha=0$ it represents solid body rotation: $\textbf{u}=(\beta y, -\beta x, 0)$. A spherical particle released in the plane $z=0$ will remain there due to the symmetry of the flow.
Substituting (\ref{eq-x_p}) and (\ref{eq-u}) (assuming $z=0$) in equation (\ref{em2}), and taking into account that no gravity is applied,  the Fax\'{e}n corrections are 0 for a linear velocity field and $\frac{Du}{Dt}={\mathcal{R}}\left[s^2(x+iy)\textbf{e}\right]$, yields the following quadratic relation for $s$:
\begin{eqnarray}
-\omega^2\left(m_p+\frac{1}{2}m_f\right) = 6\pi a\mu (-s+i\omega)
+\frac{3}{2}m_f s^2+c_{\textrm{B}}\sqrt{\frac{\omega\pi}{2}}(\omega+is)(1+i)~.
\end{eqnarray}
There are two solutions for $s$, but one of the solutions of $s$ results in an unphysical particle trajectory and is therefore discarded.
\section{Time dependent velocity field}
\noindent In this appendix the particle trajectory is derived given the uniform, time dependent velocity field (\ref{eq-vel_field}). The particle is released with an initial velocity $\textbf{u}_p(0)$. For a uniform velocity field Eq. (\ref{em2}) can be simplified to
\begin{multline}
-\left(m_p+\frac{1}{2}m_f\right)\frac{\textrm{d}\textbf{w}}{\textrm{dt}} = 6\pi a\mu\textbf{w}
+(m_f-m_p)\frac{\textrm{d}\textbf{u}}{\textrm{dt}} -(m_p-m_f)g\textbf{e}_z\\
+c_\textrm{B}\int_0^tK_{\textrm{B}}(t-\tau)\frac{\textrm{d} \textbf{w}(\tau)}{\textrm{d}\tau}\textrm{d}\tau,\label{eq-sim}
\end{multline}
where $\textbf{w}=\textbf{u}-\textbf{u}_p$. The velocity field $\textbf{u}$ will be expanded in a Fourier series, $\textbf{u}(t)=\sum_{n=-\infty}^\infty \textbf{u}_ne^{in\omega t}$. The Laplace transform of $\textbf{w}$ is given by $\textbf{W}(s)=\int_{0}^\infty e^{-st}\textbf{w}(t)\textrm{d}t$, and the Laplace transform of equation (\ref{eq-sim}) reads
\begin{multline}
-\left(m_p+\frac{1}{2}m_f+c_\textrm{B}\sqrt{\frac{\pi}{s}}\right)\left(s\textbf{W}-\textbf{w}(0)\right) = 6\pi a\mu\textbf{W}-\frac{(m_p-m_f)}{s}g\textbf{e}_z\\+(m_f-m_p)\sum_{n=-\infty}^\infty\textbf{u}_n \frac{in\omega}{s-in\omega}~.
\end{multline}
Using spitting in partial fractions this yields for $\textbf{W}(s)$
\begin{multline}
\textbf{W}(s)=\frac{c}{\sqrt{s}}\left(\frac{c_+}{\sqrt{s}+c_-}-\frac{c_-}{\sqrt{s}+c_+}\right)\left(\textbf{w}(0)-\frac{m_p-m_f}{6\pi a\mu}g\textbf{e}_z\right)+\frac{m_p-m_f}{6\pi a\mu}\frac{g}{s}\textbf{e}_z\\
+\sum_{n=-\infty}^\infty\textbf{c}_n\left\{\frac{1}{(c_++\sqrt{in\omega})(c_-+\sqrt{in\omega})(s-in\omega)}
+\frac{\sqrt{in\omega}(c_++c_-)}{(c_+^2-in\omega)(c_-^2-in\omega)\sqrt{s}(\sqrt{s}+\sqrt{in\omega})}\right.\\
\left.+\frac{c}{\sqrt{s}}\left(\frac{c_+}{(c_+^2-in\omega)(\sqrt{s}+c_+)}-\frac{c_-}{(c_-^2-in\omega)(\sqrt{s}+c_-)}\right)\right\},
\end{multline}
with $c_+$, $c_-$, $c$ and $\textbf{c}_n$ constants given by
\begin{eqnarray}
c_\pm=\frac{c_\textrm{B}\sqrt{\pi}\pm\sqrt{c_\textrm{B}^2\pi-12\pi a\mu(2m_p+m_f)}}{2m_p+m_f}~~,\nonumber\\
c=\frac{2m_p+m_f}{2\sqrt{c_\textrm{B}^2\pi-12\pi a\mu(2m_p+m_f)}}~~,~~~~~~~
\textbf{c}_n=\frac{in\omega(m_p-m_f)\textbf{u}_n }{m_p+\frac{1}{2}m_f}.
\end{eqnarray}
Transformation back to physical space results in
\begin{multline}
\textbf{w}(t)=c\left[c_+\psi\left(c_-\sqrt{t}\right)-c_-\psi\left(c_+\sqrt{t}\right)\right]\left(\textbf{w}(0)-\frac{m_p-m_f}{6\pi a\mu}g\textbf{e}_z\right) +\frac{m_p-m_f}{6\pi a\mu}g\textbf{e}_z\\
+\sum_{n=-\infty}^\infty\textbf{c}_n\left\{\frac{1}{(c_++\sqrt{in\omega})(c_-+\sqrt{in\omega})}e^{in\omega t}
+\frac{\sqrt{in\omega}(c_++c_-)}{(c_+^2-in\omega)(c_-^2-in\omega)}\psi\left(\sqrt{in\omega t}\right)\right.\\
\left.
+\frac{cc_+}{c_+^2-in\omega}\psi\left(c_+\sqrt{t}\right)-\frac{cc_-}{c_-^2-in\omega}\psi\left(c_-\sqrt{t}\right)\right\},
\end{multline}
with $\psi(z)=\exp\left(z^2\right)\textrm{erfc}\left(z\right)$.
\bibliography{bibliography}
        \bibliographystyle{unsrt}

\end{document}